\documentclass[aps,prl,showpacs,twocolumn,floats,epsfig]{revtex4}
\usepackage{amssymb}
\usepackage{amsbsy}
\usepackage{amsmath}
\usepackage{epsfig}
\usepackage{amsfonts}
\usepackage{bm}
\usepackage{graphicx}
\usepackage[breaklinks,colorlinks = true,linkcolor =
blue,urlcolor = blue,citecolor = blue,anchorcolor =
green,bookmarks=true]{hyperref}
\usepackage{mathrsfs}
\allowdisplaybreaks

\newcommand{\gae}{\lower 2pt \hbox{$\,
\buildrel{\scriptstyle >}\over {\scriptstyle \sim}\,$}}
\newcommand{\lae}{\lower 2pt \hbox{$\,
\buildrel{\scriptstyle <}\over {\scriptstyle \sim}\,$}}
\begin{document}

\title {Non-equilibrium dynamic critical scaling of the
quantum Ising chain}

\author{Michael Kolodrubetz$^{1}$, Bryan K. Clark$^{1,2}$,
David A. Huse$^{1,2}$}

\affiliation{ $^{1}$Department of Physics, Princeton
University, Princeton, NJ 08544, USA. $^{2}$Princeton Center
for Theoretical Science.}

\date{\today}

\begin{abstract}

We solve for the time-dependent finite-size scaling
functions of the 1D transverse-field Ising chain during a
linear-in-time ramp of the field through the quantum critical point.
We then simulate Mott-insulating bosons in a tilted
potential, an experimentally-studied system in the same
equilibrium universality class,
and demonstrate that universality
holds for the dynamics as well.
We find qualitatively athermal features of
the scaling functions, such as negative spin correlations,
and show that they should be
robustly observable within present cold atom experiments.

\end{abstract}


\maketitle

The Kibble-Zurek (KZ) mechanism describes the dynamics of a
system as it is ramped across a phase transition at a finite
rate. Kibble first introduced this idea to model symmetry
breaking during cooling of the early universe
\cite{Kibble1976_1}. Later, Zurek showed that condensed
matter systems exhibit similar behavior in the context of
slowly ramping across the superfluid phase transition in
liquid $^4 $He \cite{Zurek1985_1}. Both proposals looked at
ramps from deep in the disordered phase to deep in the
ordered phase. Historically, this has been the primary focus
of research on the KZ mechanism (c.f.e.
\cite{Damski2005_1,DelCampo2010_1}), although recent work
has explored ramps that end at the critical point
\cite{DeGrandi2011_1, DeGrandi2010_1,Kolodrubetz2012_1}.

However, the scaling theory of the KZ mechanism 
applies more generally than these two limits. Near the
quantum critical point (QCP), observables measured during
the ramp are postulated to exhibit universal scaling forms
\cite{Erez2012_1, Deng2008_1}. In this Letter we solve for
the scaling functions of excess heat and equal-time spin
correlations in the 1D transverse-field Ising (TFI) chain
for two scaling directions -- time $t$ during the ramp and
finite size $L$ -- and investigate the qualitative features
of these observables. In many cases,
the quantum dynamics of a closed non-integrable quantum
system lead to a thermal state. In contrast, we show
here that the spin correlation functions are qualitatively
different from those of any thermal state. 
We also find a ramp protocol where the state at long time
not only does not thermalize, but also does not dephase to
the diagonal (``generalized Gibbs'') ensemble.
We provide evidence for the universality of
the dynamics by using time-dependent matrix product states
(tMPS) to simulate the experimentally-realizable,
non-integrable model of Mott insulating bosons in a tilted
potential \cite{Sachdev2002_1,Simon2011_1}. We
see that the athermal features of the TFI chain are
robust for small systems and discuss the
prospects of seeing scaling
collapse in present-day experiments.

\emph{Transverse-field Ising chain - } The Hamiltonian for
the TFI chain on an $L$-site 1D lattice is \begin{equation}
H=-J \displaystyle\sum_{j=1}^{L} \left[
\left(1-\lambda\right) s^x_j + s^z_j s^z_{j+1} \right],
\end{equation} where $\lambda$ is a tunable transverse
magnetic field and $s^{x,z}$ are Pauli matrices with
periodic boundary conditions ($s_1=s_{L+1}$). We work in
units where $\hbar=1$ and $J=1/2$. This model has a
quantum phase transition (QPT) at $\lambda_c=0$
\cite{Sachdev1999_1}. In equilibrium, the system is in a
disordered paramagnetic (PM) phase for $\lambda < 0$, while
for $0<\lambda<2$, the system is in a ferromagnetic (FM)
phase with two degenerate ordered ground states in the limit
$L\rightarrow\infty$.

The Hamiltonian of the TFI chain can be diagonalized by
applying a Jordan-Wigner transformation
\cite{Sachdev1999_1,Dziarmaga2005_1}: \begin{equation} s^x_j
= 1-2c^\dagger_j c_j, \hspace{10pt} s^z_j = -(c_j +
c^\dagger_j) \displaystyle \prod_{m < j} s^x_m,
\end{equation} where $c^\dagger_j$ creates a fermion at site
$j$. As the Hamiltonian conserves fermion number modulo 2
and the ground state has an even number of fermions
\cite{Dziarmaga2005_1}, we restrict ourselves to that
subspace. Fourier transforming, \begin{equation} H=\sum_k
\left[ (1-\lambda-\cos k) c_k^\dagger c_k + \frac{\sin k}{2}
(c_k^\dagger c_{-k}^\dagger + c_{-k} c_k) \right].
\end{equation} This Hamiltonian can only excite fermions in
momentum pairs $(k,-k)$, so we introduce a pseudo-spin
$\sigma$ corresponding to whether the $(k,-k)$ pair is
filled ($\sigma^z=1$) or unfilled ($\sigma^z=-1$). Then in
the sector where all fermions are paired, which contains the
ground state, $H=\sum_{k=0}^\pi H_k$, where 
\begin{equation}
H_k=(1-\lambda-\cos k) \sigma^z + (\sin k) \sigma^x ~.
\end{equation} Note that $H$ is a free-fermion Hamiltonian,
so the TFI chain is integrable.

In a KZ ramp, $\lambda$ is varied as a function of time near
the critical point. For simplicity, we focus primarily on
the case of a linear ramp, $\lambda(t)=vt$, where $v$ is the
ramp rate and the QCP is at $\lambda=t=0$. The ramp begins
at $t=-\infty$ deep in the disordered phase, where the
wavefunction is initialized in the ground state of the
instantaneous Hamiltonian $H(t=-\infty)$. For an infinite
system, the Hamiltonian is gapless at the critical point.
Therefore, it is impossible to ramp slowly enough through
the QCP to avoid creating excitations and to produce true
long range order in the ordered phase.

Near the QCP, critical slowing down tells us that the
characteristic time scale $\xi_t \sim \lambda^{-\nu z}$
becomes arbitrarily large, where $\nu$ and $z$ are the
(positive) correlation length and dynamic critical exponents
respectively. Thus for a non-zero ramp rate, there exists a
Kibble time, $-t_K$, at which the lowest momentum modes stop
following the ramp adiabatically and become significantly
excited. More explicitly, 
\begin{eqnarray} 
|t_K|&=&\xi_t
(\lambda(|t_K|) ) \sim |v t_K|^{-\nu z} \nonumber \\
\Rightarrow t_K &\sim& v^{-\nu z/(1+ \nu z)} ~.
\label{eq:tK} 
\end{eqnarray}
For the TFI chain $\nu=z=1$, so
$t_K \sim v^{-1/2}$. One can similarly define a length
$\ell_K \sim t_K^{1/z}$ such that $t_K$ and $\ell_K$ set the
characteristic time and length scales for the Kibble-Zurek
scaling forms \cite{Erez2012_1, Deng2008_1}.

For a KZ ramp of the TFI chain, the wavefunction
$|\psi(t)\rangle$ can be written as a product $| \psi(t)
\rangle=\otimes_k |\Psi_k (t)\rangle$, where each mode
evolves independently as 
\begin{eqnarray}
\nonumber 
i d |\Psi_k\rangle/d t &=& H_k(t) |\Psi_k\rangle\\
 &=& \left[(1-v t - \cos k) \sigma_k^z + (\sin k) \sigma_k^x
\right] |\Psi_k\rangle ~. 
\label{eq:schr_k} 
\end{eqnarray}
The KZ scaling limit is defined as taking $v\to0$ with
$\tau$, $\Lambda$, and $\kappa$ constant \cite{Erez2012_1,
Deng2008_1}, where 
\begin{eqnarray}
\nonumber 
\tau & \equiv & t v^{1/2} \sim t/t_K \\
\nonumber
\Lambda & \equiv & L
v^{1/2} \sim L/\ell_K\\
\nonumber
\kappa & \equiv & k
v^{-1/2} \sim k \ell_K \\
&=&\frac{2\pi(n+1/2)}{\Lambda}.
\label{eq:scvar}
\end{eqnarray}
Here $n = 0,1,..., (L/2)-1$
indexes the modes. Note that $L \to \infty$ in the scaling
limit, as $\ell_K \to \infty$ and $\Lambda$ is constant. In
this KZ scaling limit, eq. \ref{eq:schr_k} simplifies to
\begin{equation}
\label{eq:velexp}
i d\Psi_\kappa/d\tau=
\left( -\tau \sigma_\kappa^z + \kappa \sigma_\kappa^x
\right) \Psi_\kappa ~.
\end{equation}
Each mode
$\Psi_\kappa(\tau)$ in (\ref{eq:velexp}) has the form of a
Landau-Zener (LZ) equation, which we solve in terms of parabolic cylinder
functions \cite{Zener1932_1}. Note that $\Psi_\kappa(\tau)$
is expressed solely in terms of the scaled variables $\tau$
and $\kappa$.

\begin{figure*}
\includegraphics[width=\linewidth]{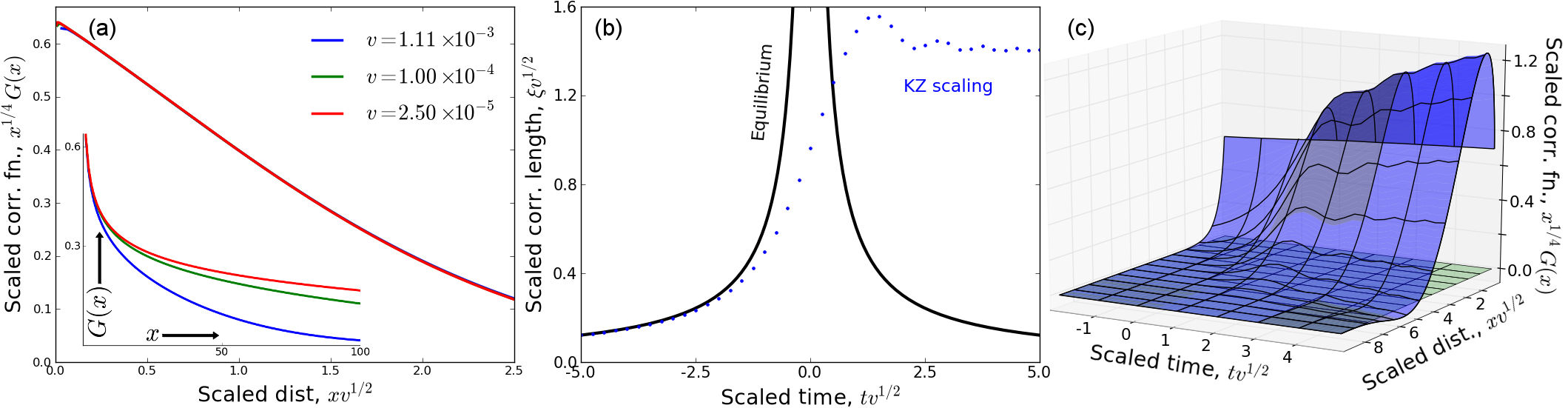}
\caption{(color online) 
(a) Spin-spin correlation function
for the TFI chain for a ramp to the QCP ($t=0$), showing
scaling collapse for a wide range of slow ramp rates. The
inset shows correlation functions prior to scaling. (b)
Scaled correlation length and (c) correlation function as a
function of scaled time (b,c) and distance (c) during the
ramp. The KZ correlation length deviates from equilibrium
near $tv^{1/2}=-2.5$ and remains finite at the QCP and
beyond. We define the correlation length as $\xi v^{1/2} =
\sqrt{\int_0^\infty g(\chi) \chi^2 d\chi / \int_0^\infty
g(\chi) d\chi}$, where $\chi = xv^{1/2}$. All data in this
figure are in the KZ thermodynamic limit, $Lv^{1/2} \gg 1$
(see text).}
\label{fig:ising}
\end{figure*}

\emph{Observables - } A crude measure of deviation from
adiabaticity is the excess heat, defined as
\begin{equation}
\label{eq:Qgeneral}
\begin{split} Q(t)=\sum_{\kappa} \left[
\right. & \langle \Psi_\kappa(t) | H_\kappa(t) | \Psi_\kappa
(t) \rangle - \\
& \left. \langle \Psi^0_\kappa (t) |
H_\kappa(t) | \Psi^0_\kappa (t) \rangle \right], \end{split}
\end{equation}
where $| \Psi^0_\kappa (t) \rangle$ is the
instantaneous ground state of $H_\kappa (t)$. Defining
excited state occupancy $p_\kappa^\mathrm{exc}=1-\left|
\langle \Psi_\kappa^0 | \Psi_\kappa \rangle\right|^2$ and
energy $E_{\kappa}^\mathrm{exc}=2 v^{1/2} \sqrt{\tau^2 +
\kappa^2}$, the scaled excess heat density is
\begin{eqnarray}
\nonumber
q(\tau,\Lambda) & \equiv &
\frac{Q(t)}{vL}\\
\label{eq:q}
& = & \frac{2}{\Lambda}
\sum_\kappa p_\kappa^\mathrm{exc}(\tau,\Lambda) \sqrt{\tau^2
+ \kappa^2}~.
\end{eqnarray}

For the equal-time $s^z$-$s^z$ correlation function, we
postulate a non-equilibrium scaling form, and check for
scaling collapse. The correlation function is defined as
$G(x)=\langle s^z_j s^z_{j+x}\rangle$, assuming translation
invariance. In equilibrium, $s^z$ has scaling dimension 1/8,
so the scaled correlation function is
\begin{equation}
g(\tau,\Lambda,\chi) \equiv G(x,t,L,v) x^{1/4},
\end{equation}
where $\chi \equiv xv^{1/2}$. We break up
each site into a pair of Majorana fermions
\cite{Sachdev1999_1}, such that the correlation function is
the Pfaffian of a matrix whose elements are pairwise
expectation values of the Majoranas \cite{Barouch1971_1}. We
evaluate the Pfaffian numerically \cite{FootnoteToeplitz_1},
and find that good scaling collapse occurs for small $v$
(see fig. \ref{fig:ising}a).

\begin{figure*}
\includegraphics[width=\linewidth]{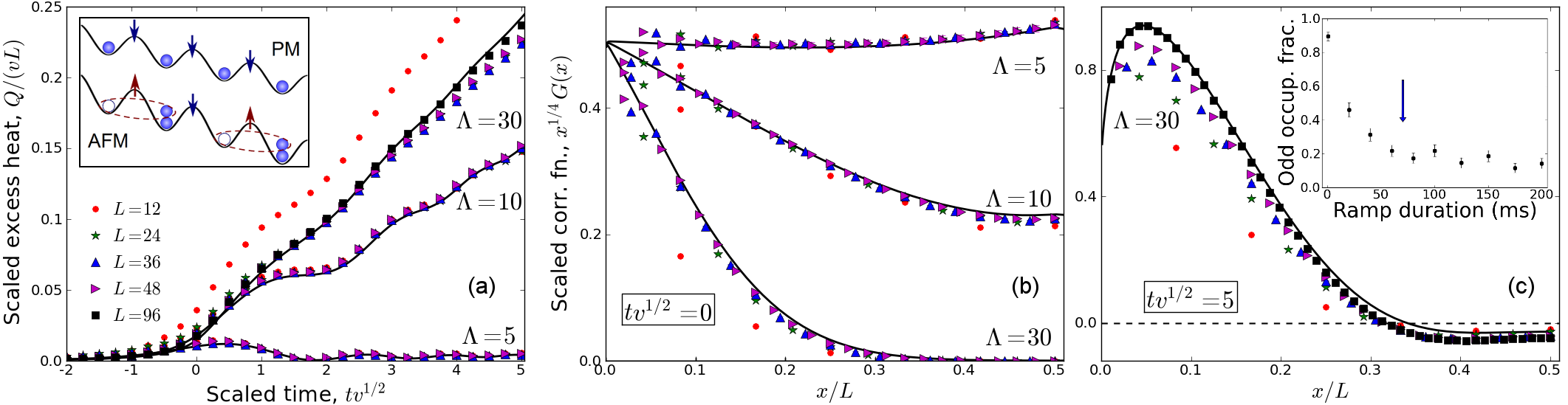}
\caption{(color online) Results of tMPS simulations for
ramping the experimentally-realizable MITP model (see text),
which is illustrated in the inset to (a). The data show
scaling collapse of MITP (colored dots) to the scaling limit
of the TFI chain (black lines) as a function of scaled time
$\tau=t v^{1/2}$ (a), scaled system size $\Lambda=Lv^{1/2}$
(a,b), and scaled length $x/L$ (b,c)
\cite{Footnote_ScalingConversion}. We have checked that data 
for $\Lambda=30$ is near the
$\Lambda=\infty$ limit for the TFI model (not shown), although finite size effects
can still be seen in (c). The arrow in the inset to (c) indicates the maximum 
ramp duration that produces athermal negative correlations
in a twelve-site bosonic chain
(data from Ref.~\onlinecite{Simon2011_1}, see
\cite{FootnoteBHE_1} for details).}
\label{fig:bhe}
\end{figure*}

During the initial part of the ramp ($\tau \ll -1$), the
system is very nearly in equilibrium (figs. \ref{fig:bhe}a,
\ref{fig:ising}b). Deep on the FM side of the ramp
($\tau\gg1$), LZ physics tells us that the
excitation probability of mode $\kappa$ is
\cite{Zener1932_1}
\begin{equation}
p^\mathrm{exc}_{\kappa}=\exp(-\pi \kappa^2) ~.
\label{eq:lz}
\end{equation}
In between, at finite positive
$\tau$, the excitation probability for each mode -- given by
the LZ equation -- oscillates as a function of scaled time
before converging to (\ref{eq:lz}). These oscillations show
up in the excess heat (fig. \ref{fig:bhe}a) and the
correlation function (fig. \ref{fig:ising}b,c).

Finite-size effects enter by opening up a gap at the QCP. In
the regime of small scaled size $\Lambda \lesssim 1$, the
finite-size gap dominates, the system remains near the
ground state, and the solution can be understood
perturbatively. In the limit $\Lambda \to \infty$, on the
other hand, finite-size effects vanish; we will refer to
this as the thermodynamic limit of KZ scaling (KZ-TDL).
Looking at the correlations right at the QCP (fig.
\ref{fig:bhe}b), we see the finite-size crossover from
near-equilibrium power-law correlations for $\Lambda=5$ to a
non-equilibrium exponential decay of correlations for
$\Lambda=30$. Similarly, in fig. \ref{fig:bhe}a we see that
the scaled excess heat $q$ crosses over from being small at
$\Lambda=5$, where the system is nearly adiabatic, to
much larger values at larger $\Lambda$.

The TFI chain is integrable, so although a KZ ramp through
the critical point creates excitations, the resulting
excited states differ markedly from equilibrium thermal
states at the same energies. From (\ref{eq:lz}), we see that
at long times the populations of modes with $\kappa^2 <
(\log 2) / \pi$ are inverted, i.e. at a negative effective
temperature. This leads to qualitatively athermal physics in
the scaled correlation function, which goes substantially
negative by $\tau=5$ over a range of scaled distances, as
can be seen in figs. \ref{fig:bhe}c and \ref{fig:ising}c.
This is qualitatively different from any thermal state,
which would have a finite correlation length but never
negative correlations. Similar behavior has been seen
outside the scaling regime for both slow \cite{Cincio2007_1}
and instantaneous \cite{Sengupta2004_1} quenches of the TFI
chain and the XY model \cite{Cherng2006_1}.

Finally, we examine what happens formally as $\tau \to
\infty$ while remaining in the scaling limit; in particular,
does the system dephase? For comparison, if one were to stop
the ramp at some value $\lambda_f$ and wait a very long
time, the phase differences between modes would
increase to the point where the phases are essentially random;
this process is known as dephasing.  Dephasing
in integrable systems is a well-studied problem, and
it has been shown that, in the long time limit, the
observables of the time-evolved pure state
approach those of the
generalized Gibbs ensemble (GGE, see ref.
\cite{Rigol2007_1}), given by removing all phase
information from each mode.   
Here we define the GGE at time $\tau$ as the
dephased ensemble that one would approach upon
freezing the current Hamiltonian and waiting a long time,
\begin{equation}
\begin{split}
\rho_\mathrm{GGE} (\tau) = \prod_\kappa \Big[
& (1-p^\mathrm{exc}_\kappa (\tau) ) | \Psi^0_\kappa (\tau) \rangle 
\langle \Psi^0_\kappa (\tau) | + \\
& p^\mathrm{exc}_\kappa (\tau)
| \Psi^1_\kappa (\tau) \rangle 
\langle \Psi^1_\kappa (\tau) |
\Big ] ~,
\end{split}
\end{equation}
where $| \Psi^1_\kappa (\tau) \rangle$ is the excited
state of mode Hamiltonian $H_\kappa$. 
In the limit $\tau \to \infty$, the mode Hamiltonians
asymptote to $H_\kappa \propto -\sigma^z_\kappa$,
so the GGE approaches a fixed
value with $|\Psi^0_\kappa \rangle \to |\uparrow
\rangle$, $|\Psi^1_\kappa \rangle \to |\downarrow \rangle$, and
$p^\mathrm{exc}_\kappa \to e^{-\pi \kappa^2}$.

To see if $\tau \to \infty$ leads to dephasing,
we consider the phase difference $\Delta \varphi$ between
characteristic modes $\kappa=0$ and $\kappa=1$, since
excitations are exponentially suppressed for $\kappa \gae
1$. Starting from some time $\tau_i \gg 1$, after which the
dynamics is effectively adiabatic, the energy difference
$\Delta E$ and phase difference $\Delta \varphi$ are
\begin{eqnarray}
\nonumber \Delta E (\tau) = \sqrt{\tau^2 +
1} - \tau \approx \frac{1}{2\tau} \\
\Delta \varphi =
\int_{\tau_i}^{\tau_f} \Delta E(\tau) d\tau \approx
\frac{1}{2} \log (\tau_f/\tau_i)
\end{eqnarray}
Since $\Delta \varphi \to \infty$ as $\tau_f \to \infty$, 
the phase information between modes is lost in the long
time limit, so the observables
approach those of the GGE.

We note that, for non-linear ramps (say cubic
ramps, $\lambda \sim t^3$), a similar scaling
theory can be written down \cite{Erez2012_1}.  
Then the above argument again holds, except now
\begin{equation}
\Delta \varphi \approx
\int_{\tau_i}^{\tau_f} \frac{1}{2\tau^3} \to \mathrm{const.}
\end{equation}
as $\tau_f \to \infty$, implying that phase information
remains, and the cubic ramp does not approach the GGE
\cite{Footnote_CubicFeasibility}.  To summarize,
both linear and cubic ramps exhibit athermal behavior, 
such as negative correlations, which come from the inversion
of low-momentum modes, but only the linear ramp dephases
to the GGE in the long time limit.

\emph{Tilted bosons - } While the TFI chain is a beautiful
theoretical model, it is difficult to realize in the lab.
However, there have been a number of recent experimental
advances with other, non-integrable models in the Ising
universality class \cite{Zhang2009_1,Simon2011_1,
Kim2011_1}. Here we
focus on one, the Mott insulator in a tilted potential
(MITP), realized experimentally in ref. \cite{Simon2011_1}.
The MITP consists of a one-dimensional lattice containing a
Mott insulator with one boson per site. By adding a
sufficiently large potential gradient (tilt) $\delta$, the
system undergoes a QPT from 1 boson per site to alternately
0 and 2 per site, creating dipoles on every other bond (see
fig. \ref{fig:bhe}a inset). This QCP is in the Ising
universality class \cite{Sachdev2002_1}, and can be
described by an effective spin Hamiltonian
\begin{equation}
\label{eq:Hmitp}
H_\mathrm{eff}= \mathcal{P} \left\{ u
\sum_l \left[ \delta \, \frac{S^z_l+1}{2} - S^x_l \right] \right\}
\mathcal{P},
\end{equation}
where $l$ labels the bonds,
$S^{x,z}$ are Pauli matrices residing on the bonds,
and $\mathcal{P}$ is a projector implementing the constraint
that no two neighboring bonds are both spin up. The energy
scale is set by $u=\sqrt{2} w$, where $w$ is the hopping
energy of the lattice bosons; we work in units with $u=1$.
In analogy with the TFI chain, we define the parameters
$\lambda=\delta_c-\delta$ and $v=\partial_t
\lambda$ \cite{Footnote_ScalingConversion}, where $\delta_c
\cong -1.31$ is the location of the QCP
\cite{Sachdev2002_1}. The correlation function is given by
$G(x)=(-1)^x \left( \langle S^z_j S^z_{j+x} \rangle -
\langle S^z_j \rangle \langle S^z_{j+x} \rangle \right)$.

We simulate a linear ramp via tMPS, as described in ref.
\cite{Kolodrubetz2012_1}. Fig. \ref{fig:bhe} shows
calculated observables of the MITP model compared to the
Ising scaling forms derived earlier. We see clear scaling
collapse in all three scaling directions: ramp time $t$,
system size $L$, and distance $x$. As expected, simulations
do not collapse as well for smaller system sizes. While the
dynamic range of the simulations is limited, and the scaling
collapse for finite systems is imperfect, we consider our
results to be strong numerical evidence for the postulated
universality of KZ scaling for systems in the Ising
universality class.

We note that, while scaling collapse for the system sizes
considered is not perfect at $tv^{1/2}=5$, the correlation
function goes negative as predicted from the TFI scaling
function. Therefore, the athermal physics of the KZ ramp is
robust against small system sizes and the breaking of
integrability, and is a qualitative feature of our model
that should be visible experimentally. A major open question
is whether such athermal behavior will manifest in KZ ramps
near non-integrable QCPs, where the relatively short-time
dynamics of the KZ ramp will be in competition with the
long-time expectation of eigenstate thermalization
\cite{Rigol2008_1}.

Finally, we compare our time scales to the those of the real
experimental system \cite{Simon2011_1}. Using
the experimental parameters
$w\approx 10$ Hz and $U\approx400$ Hz, where $U$ is the
on-site repulsion, we estimate
the ramp rate necessary to see athermal negative
spin correlation; our predicted ramp
rates are shown in the inset to Fig. \ref{fig:bhe}c compared
to experimental data from Ref.~\onlinecite{Simon2011_1}.
Clearly the necessary ramp rates are well
within the experimental range \cite{FootnoteBHE_1}.

In conclusion, we have solved for the full Kibble-Zurek
dynamic scaling forms of the one-dimensional transverse
field Ising chain at zero temperature. We provided numerical
evidence for the universality of these scaling relations via
tMPS simulations of the MITP model. Finally, we determined
that the relevant time scales for seeing these effects
experimentally are within the reach of current technology. To
see full scaling collapse the experimental system sizes need
to be larger, but they should already be sufficient to see
qualitatively similar athermal physics.

\emph{Acknowledgments -- }
We would like to thank S. Sondhi, A. Polkovnikov, 
K. Sengupta, D. Pekker, A.
Chandran, and A. Erez for valuable discussions. This work
was supported in part by ARO Award W911NF-07-1-0464 with
funds from the DARPA OLE Program. Some of the computation
was performed using the Extreme Science and Engineering
Discovery Environment (XSEDE), which is supported by
National Science Foundation grant number OCI-1053575.
Additional computation was done on the Feynman and
Della clusters at Princeton.

\bibliography{/home/mkolodru/Dropbox/References/References}

\begin{thebibliography}{28}
\expandafter\ifx\csname natexlab\endcsname\relax\def\natexlab#1{#1}\fi
\expandafter\ifx\csname bibnamefont\endcsname\relax
  \def\bibnamefont#1{#1}\fi
\expandafter\ifx\csname bibfnamefont\endcsname\relax
  \def\bibfnamefont#1{#1}\fi
\expandafter\ifx\csname citenamefont\endcsname\relax
  \def\citenamefont#1{#1}\fi
\expandafter\ifx\csname url\endcsname\relax
  \def\url#1{\texttt{#1}}\fi
\expandafter\ifx\csname urlprefix\endcsname\relax\def\urlprefix{URL }\fi
\providecommand{\bibinfo}[2]{#2}
\providecommand{\eprint}[2][]{\url{#2}}

\bibitem[{\citenamefont{Kibble}(1976)}]{Kibble1976_1}
\bibinfo{author}{\bibfnamefont{T.}~\bibnamefont{Kibble}}, \bibinfo{journal}{J
  Phys. A: Math. Gen.} \textbf{\bibinfo{volume}{9}}, \bibinfo{pages}{1387}
  (\bibinfo{year}{1976}).

\bibitem[{\citenamefont{Zurek}(1985)}]{Zurek1985_1}
\bibinfo{author}{\bibfnamefont{W.~H.} \bibnamefont{Zurek}},
  \bibinfo{journal}{Nature} \textbf{\bibinfo{volume}{317}},
  \bibinfo{pages}{505} (\bibinfo{year}{1985}).

\bibitem[{\citenamefont{Damski}(2005)}]{Damski2005_1}
\bibinfo{author}{\bibfnamefont{B.}~\bibnamefont{Damski}},
  \bibinfo{journal}{Phys. Rev. Lett.} \textbf{\bibinfo{volume}{95}},
  \bibinfo{pages}{035701} (\bibinfo{year}{2005}).

\bibitem[{\citenamefont{del Campo et~al.}(2010)\citenamefont{del Campo,
  De~Chiara, Morigi, Plenio, and Retzker}}]{DelCampo2010_1}
\bibinfo{author}{\bibfnamefont{A.}~\bibnamefont{del Campo}},
  \bibinfo{author}{\bibfnamefont{G.}~\bibnamefont{De~Chiara}},
  \bibinfo{author}{\bibfnamefont{G.}~\bibnamefont{Morigi}},
  \bibinfo{author}{\bibfnamefont{M.~B.} \bibnamefont{Plenio}},
  \bibnamefont{and} \bibinfo{author}{\bibfnamefont{A.}~\bibnamefont{Retzker}},
  \bibinfo{journal}{Phys. Rev. Lett.} \textbf{\bibinfo{volume}{105}},
  \bibinfo{pages}{075701} (\bibinfo{year}{2010}).

\bibitem[{\citenamefont{De~Grandi et~al.}(2011)\citenamefont{De~Grandi,
  Polkovnikov, and Sandvik}}]{DeGrandi2011_1}
\bibinfo{author}{\bibfnamefont{C.}~\bibnamefont{De~Grandi}},
  \bibinfo{author}{\bibfnamefont{A.}~\bibnamefont{Polkovnikov}},
  \bibnamefont{and} \bibinfo{author}{\bibfnamefont{A.~W.}
  \bibnamefont{Sandvik}}, \bibinfo{journal}{Phys. Rev. B}
  \textbf{\bibinfo{volume}{84}}, \bibinfo{pages}{224303}
  (\bibinfo{year}{2011}).

\bibitem[{\citenamefont{De~Grandi et~al.}(2010)\citenamefont{De~Grandi,
  Gritsev, and Polkovnikov}}]{DeGrandi2010_1}
\bibinfo{author}{\bibfnamefont{C.}~\bibnamefont{De~Grandi}},
  \bibinfo{author}{\bibfnamefont{V.}~\bibnamefont{Gritsev}}, \bibnamefont{and}
  \bibinfo{author}{\bibfnamefont{A.}~\bibnamefont{Polkovnikov}},
  \bibinfo{journal}{Phys. Rev. B} \textbf{\bibinfo{volume}{81}},
  \bibinfo{pages}{012303} (\bibinfo{year}{2010}).

\bibitem[{\citenamefont{Kolodrubetz et~al.}(2012)\citenamefont{Kolodrubetz,
  Pekker, Clark, and Sengupta}}]{Kolodrubetz2012_1}
\bibinfo{author}{\bibfnamefont{M.}~\bibnamefont{Kolodrubetz}},
  \bibinfo{author}{\bibfnamefont{D.}~\bibnamefont{Pekker}},
  \bibinfo{author}{\bibfnamefont{B.~K.} \bibnamefont{Clark}}, \bibnamefont{and}
  \bibinfo{author}{\bibfnamefont{K.}~\bibnamefont{Sengupta}},
  \bibinfo{journal}{Phys. Rev. B} \textbf{\bibinfo{volume}{85}},
  \bibinfo{pages}{100505} (\bibinfo{year}{2012}).

\bibitem[{\citenamefont{Chandran et~al.}(2012)\citenamefont{Chandran, Erez,
  Gubser, and Sondhi}}]{Erez2012_1}
\bibinfo{author}{\bibfnamefont{A.}~\bibnamefont{Chandran}},
  \bibinfo{author}{\bibfnamefont{A.}~\bibnamefont{Erez}},
  \bibinfo{author}{\bibfnamefont{S.~S.} \bibnamefont{Gubser}},
  \bibnamefont{and} \bibinfo{author}{\bibfnamefont{S.~L.}
  \bibnamefont{Sondhi}}, \bibinfo{journal}{arXiv:1202.5277}
  (\bibinfo{year}{2012}).

\bibitem[{\citenamefont{Deng et~al.}(2008)\citenamefont{Deng, Ortiz, and
  Viola}}]{Deng2008_1}
\bibinfo{author}{\bibfnamefont{S.}~\bibnamefont{Deng}},
  \bibinfo{author}{\bibfnamefont{G.}~\bibnamefont{Ortiz}}, \bibnamefont{and}
  \bibinfo{author}{\bibfnamefont{L.}~\bibnamefont{Viola}},
  \bibinfo{journal}{EPL (Europhysics Letters)} \textbf{\bibinfo{volume}{84}},
  \bibinfo{pages}{67008} (\bibinfo{year}{2008}), ISSN
  \bibinfo{issn}{0295-5075}.

\bibitem[{\citenamefont{Sachdev et~al.}(2002)\citenamefont{Sachdev, Sengupta,
  and Girvin}}]{Sachdev2002_1}
\bibinfo{author}{\bibfnamefont{S.}~\bibnamefont{Sachdev}},
  \bibinfo{author}{\bibfnamefont{K.}~\bibnamefont{Sengupta}}, \bibnamefont{and}
  \bibinfo{author}{\bibfnamefont{S.~M.} \bibnamefont{Girvin}},
  \bibinfo{journal}{Phys. Rev. B} \textbf{\bibinfo{volume}{66}},
  \bibinfo{pages}{075128} (\bibinfo{year}{2002}).

\bibitem[{\citenamefont{Simon et~al.}(2011)\citenamefont{Simon, Bakr, Ma, Tai,
  Preiss, and Greiner}}]{Simon2011_1}
\bibinfo{author}{\bibfnamefont{J.}~\bibnamefont{Simon}},
  \bibinfo{author}{\bibfnamefont{W.~S.} \bibnamefont{Bakr}},
  \bibinfo{author}{\bibfnamefont{R.}~\bibnamefont{Ma}},
  \bibinfo{author}{\bibfnamefont{M.~E.} \bibnamefont{Tai}},
  \bibinfo{author}{\bibfnamefont{P.~M.} \bibnamefont{Preiss}},
  \bibnamefont{and} \bibinfo{author}{\bibfnamefont{M.}~\bibnamefont{Greiner}},
  \bibinfo{journal}{Nature} \textbf{\bibinfo{volume}{472}},
  \bibinfo{pages}{307} (\bibinfo{year}{2011}), ISSN \bibinfo{issn}{0028-0836}.

\bibitem[{\citenamefont{Sachdev}(1999)}]{Sachdev1999_1}
\bibinfo{author}{\bibfnamefont{S.}~\bibnamefont{Sachdev}},
  \emph{\bibinfo{title}{Quantum Phase Transitions}}
  (\bibinfo{publisher}{Cambridge University Press}, \bibinfo{year}{1999}).

\bibitem[{\citenamefont{Dziarmaga}(2005)}]{Dziarmaga2005_1}
\bibinfo{author}{\bibfnamefont{J.}~\bibnamefont{Dziarmaga}},
  \bibinfo{journal}{Phys. Rev. Lett.} \textbf{\bibinfo{volume}{95}},
  \bibinfo{pages}{245701} (\bibinfo{year}{2005}).

\bibitem[{\citenamefont{Zener}(1932)}]{Zener1932_1}
\bibinfo{author}{\bibfnamefont{C.}~\bibnamefont{Zener}},
  \bibinfo{journal}{Proceedings of the Royal Society of London A}
  \textbf{\bibinfo{volume}{137}}, \bibinfo{pages}{696} (\bibinfo{year}{1932}).

\bibitem[{\citenamefont{Barouch and McCoy}(1971)}]{Barouch1971_1}
\bibinfo{author}{\bibfnamefont{E.}~\bibnamefont{Barouch}} \bibnamefont{and}
  \bibinfo{author}{\bibfnamefont{B.~M.} \bibnamefont{McCoy}},
  \bibinfo{journal}{Phys. Rev. A} \textbf{\bibinfo{volume}{3}},
  \bibinfo{pages}{786} (\bibinfo{year}{1971}).

\bibitem[{Foo({\natexlab{a}})}]{FootnoteToeplitz_1}
\bibinfo{note}{We note that, in principle, the theory of Toeplitz determinants
  \cite{Wu1966_1} can be used to analytically solve for the correlation
  function in the limit $x\gg 1$.}

\bibitem[{\citenamefont{Cincio et~al.}(2007)\citenamefont{Cincio, Dziarmaga,
  Rams, and Zurek}}]{Cincio2007_1}
\bibinfo{author}{\bibfnamefont{L.}~\bibnamefont{Cincio}},
  \bibinfo{author}{\bibfnamefont{J.}~\bibnamefont{Dziarmaga}},
  \bibinfo{author}{\bibfnamefont{M.~M.} \bibnamefont{Rams}}, \bibnamefont{and}
  \bibinfo{author}{\bibfnamefont{W.~H.} \bibnamefont{Zurek}},
  \bibinfo{journal}{Phys. Rev. A} \textbf{\bibinfo{volume}{75}},
  \bibinfo{pages}{052321} (\bibinfo{year}{2007}).

\bibitem[{\citenamefont{Sengupta et~al.}(2004)\citenamefont{Sengupta, Powell,
  and Sachdev}}]{Sengupta2004_1}
\bibinfo{author}{\bibfnamefont{K.}~\bibnamefont{Sengupta}},
  \bibinfo{author}{\bibfnamefont{S.}~\bibnamefont{Powell}}, \bibnamefont{and}
  \bibinfo{author}{\bibfnamefont{S.}~\bibnamefont{Sachdev}},
  \bibinfo{journal}{Phys. Rev. A} \textbf{\bibinfo{volume}{69}},
  \bibinfo{pages}{053616} (\bibinfo{year}{2004}).

\bibitem[{\citenamefont{Cherng and Levitov}(2006)}]{Cherng2006_1}
\bibinfo{author}{\bibfnamefont{R.~W.} \bibnamefont{Cherng}} \bibnamefont{and}
  \bibinfo{author}{\bibfnamefont{L.~S.} \bibnamefont{Levitov}},
  \bibinfo{journal}{Phys. Rev. A} \textbf{\bibinfo{volume}{73}},
  \bibinfo{pages}{043614} (\bibinfo{year}{2006}).

\bibitem[{\citenamefont{Rigol et~al.}(2007)\citenamefont{Rigol, Dunjko,
  Yurovsky, and Olshanii}}]{Rigol2007_1}
\bibinfo{author}{\bibfnamefont{M.}~\bibnamefont{Rigol}},
  \bibinfo{author}{\bibfnamefont{V.}~\bibnamefont{Dunjko}},
  \bibinfo{author}{\bibfnamefont{V.}~\bibnamefont{Yurovsky}}, \bibnamefont{and}
  \bibinfo{author}{\bibfnamefont{M.}~\bibnamefont{Olshanii}},
  \bibinfo{journal}{Phys. Rev. Lett.} \textbf{\bibinfo{volume}{98}},
  \bibinfo{pages}{050405} (\bibinfo{year}{2007}).

\bibitem[{Foo({\natexlab{b}})}]{Footnote_CubicFeasibility}
\bibinfo{note}{We note two points about observing dephasing in practice. First,
  our arguments only hold when taking the long time limit while remaining near
  the QCP ($\lambda \ll 1$) to stay in the scaling regime. This limit is
  difficult to achieve, and is compounded by disorder, which becomes more
  important in the slow ramp, long time limit. Second, one must know the exact
  location of the QCP for the cubic ramp; a slight ``miss'' would allow one to
  linearize about the actual QCP. The slower the ramp, the more accurately one
  needs to know the QCP to see cubic scaling.}

\bibitem[{Foo({\natexlab{c}})}]{Footnote_ScalingConversion}
\bibinfo{note}{There is an overall non-universal scaling between the TFI and
  MITP models, given empirically by $v_\mathrm{MITP}=4.84 v_\mathrm{TFI}$,
  $G_\mathrm{MITP}=0.196 G_\mathrm{TFI}$, and $t_\mathrm{MITP}=0.557
  t_\mathrm{TFI}$. All values in fig. \ref{fig:ising} are those of the TFI
  chain, while fig. \ref{fig:bhe} gives those of the MITP model.}

\bibitem[{Foo({\natexlab{d}})}]{FootnoteBHE_1}
\bibinfo{note}{Our analysis neglects the effect of disorder and open boundary
  conditions, which will need to be included to fully describe the Harvard
  experiment. See the supplemental information for brief analysis of open
  boundary conditions.}

\bibitem[{\citenamefont{Zhang et~al.}(2009)\citenamefont{Zhang, Cucchietti,
  Chandrashekar, Laforest, Ryan, Ditty, Hubbard, Gamble, and
  Laflamme}}]{Zhang2009_1}
\bibinfo{author}{\bibfnamefont{J.}~\bibnamefont{Zhang}},
  \bibinfo{author}{\bibfnamefont{F.~M.} \bibnamefont{Cucchietti}},
  \bibinfo{author}{\bibfnamefont{C.~M.} \bibnamefont{Chandrashekar}},
  \bibinfo{author}{\bibfnamefont{M.}~\bibnamefont{Laforest}},
  \bibinfo{author}{\bibfnamefont{C.~A.} \bibnamefont{Ryan}},
  \bibinfo{author}{\bibfnamefont{M.}~\bibnamefont{Ditty}},
  \bibinfo{author}{\bibfnamefont{A.}~\bibnamefont{Hubbard}},
  \bibinfo{author}{\bibfnamefont{J.~K.} \bibnamefont{Gamble}},
  \bibnamefont{and} \bibinfo{author}{\bibfnamefont{R.}~\bibnamefont{Laflamme}},
  \bibinfo{journal}{Phys. Rev. A} \textbf{\bibinfo{volume}{79}},
  \bibinfo{pages}{012305} (\bibinfo{year}{2009}).

\bibitem[{\citenamefont{Kim et~al.}(2011)\citenamefont{Kim, Korenblit, Islam,
  Edwards, Chang, Noh, Carmichael, Lin, Duan, Wang et~al.}}]{Kim2011_1}
\bibinfo{author}{\bibfnamefont{K.}~\bibnamefont{Kim}},
  \bibinfo{author}{\bibfnamefont{S.}~\bibnamefont{Korenblit}},
  \bibinfo{author}{\bibfnamefont{R.}~\bibnamefont{Islam}},
  \bibinfo{author}{\bibfnamefont{E.~E.} \bibnamefont{Edwards}},
  \bibinfo{author}{\bibfnamefont{M.-S.} \bibnamefont{Chang}},
  \bibinfo{author}{\bibfnamefont{C.}~\bibnamefont{Noh}},
  \bibinfo{author}{\bibfnamefont{H.}~\bibnamefont{Carmichael}},
  \bibinfo{author}{\bibfnamefont{G.-D.} \bibnamefont{Lin}},
  \bibinfo{author}{\bibfnamefont{L.-M.} \bibnamefont{Duan}},
  \bibinfo{author}{\bibfnamefont{C.~C.~J.} \bibnamefont{Wang}},
  \bibnamefont{et~al.}, \bibinfo{journal}{New Journal of Physics}
  \textbf{\bibinfo{volume}{13}}, \bibinfo{pages}{105003}
  (\bibinfo{year}{2011}), ISSN \bibinfo{issn}{1367-2630}.

\bibitem[{\citenamefont{Rigol et~al.}(2008)\citenamefont{Rigol, Dunjko, and
  Olshanii}}]{Rigol2008_1}
\bibinfo{author}{\bibfnamefont{M.}~\bibnamefont{Rigol}},
  \bibinfo{author}{\bibfnamefont{V.}~\bibnamefont{Dunjko}}, \bibnamefont{and}
  \bibinfo{author}{\bibfnamefont{M.}~\bibnamefont{Olshanii}},
  \bibinfo{journal}{Nature} \textbf{\bibinfo{volume}{452}},
  \bibinfo{pages}{854} (\bibinfo{year}{2008}), ISSN \bibinfo{issn}{0028-0836}.

\bibitem[{\citenamefont{Wu}(1966)}]{Wu1966_1}
\bibinfo{author}{\bibfnamefont{T.~T.} \bibnamefont{Wu}},
  \bibinfo{journal}{Phys. Rev.} \textbf{\bibinfo{volume}{149}},
  \bibinfo{pages}{380} (\bibinfo{year}{1966}).

\bibitem[{\citenamefont{K\"uhner et~al.}(2000)\citenamefont{K\"uhner, White,
  and Monien}}]{Kuhner2000_1}
\bibinfo{author}{\bibfnamefont{T.~D.} \bibnamefont{K\"uhner}},
  \bibinfo{author}{\bibfnamefont{S.~R.} \bibnamefont{White}}, \bibnamefont{and}
  \bibinfo{author}{\bibfnamefont{H.}~\bibnamefont{Monien}},
  \bibinfo{journal}{Phys. Rev. B} \textbf{\bibinfo{volume}{61}},
  \bibinfo{pages}{12474} (\bibinfo{year}{2000}).

\end{thebibliography}

\appendix
\section{Supplementary information for ``Non-equilibrium dynamic critical scaling of the
quantum Ising chain''}

\begin{figure*} \includegraphics[width=.7
\linewidth]{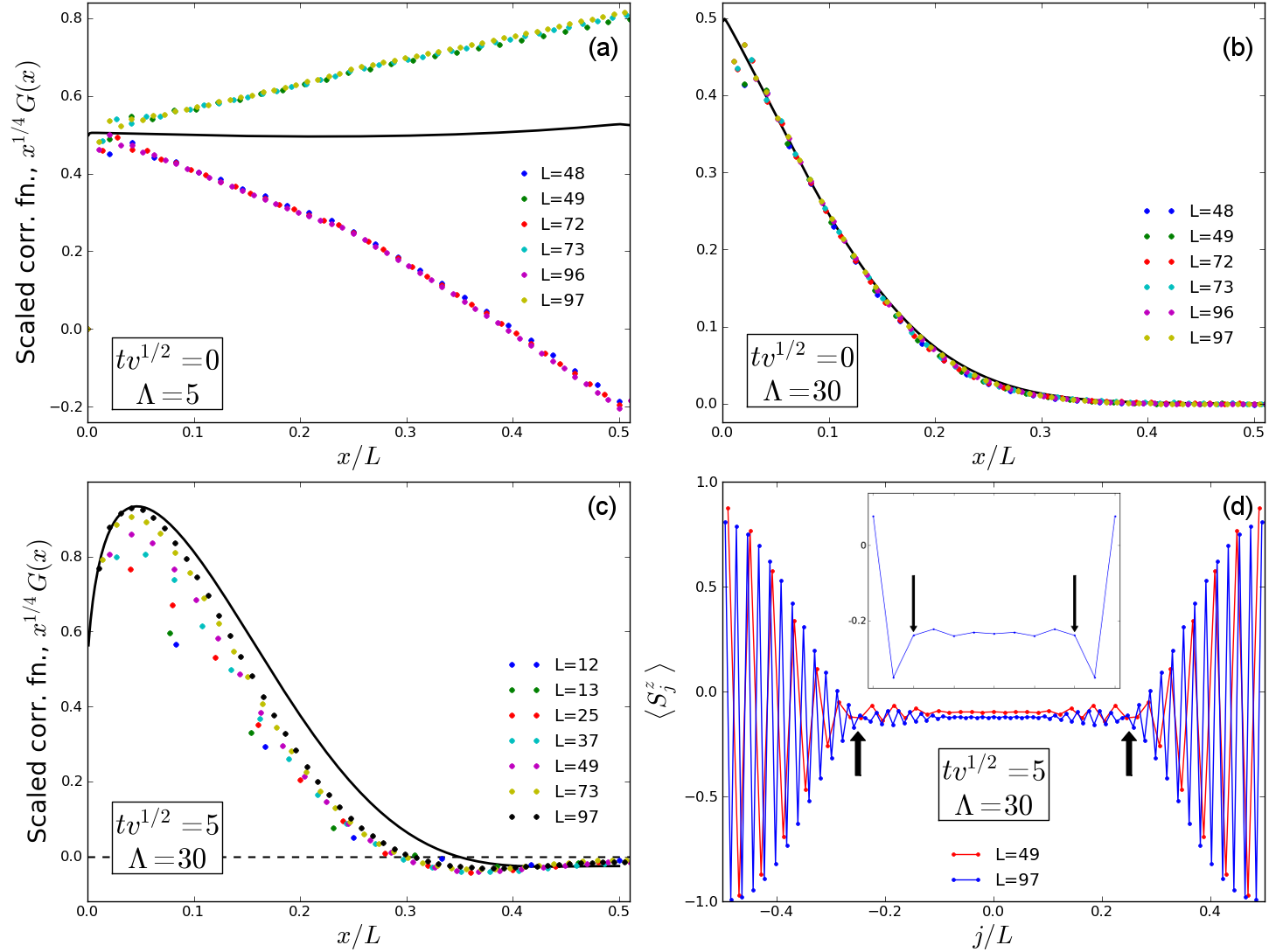} \caption{(color online)
(a-c) Spin-spin correlation function of the MITP model with
open boundary conditions, compared to the TFI scaling
function with p.b.c. (black lines).
(d) $z$ components of the
individual spins, $S_j^z$, for the same parameters as (c)
with $L=$ 97, 49, and 13 (inset) sites, where $j$ is the
site index. Arrows indicate the $180^\circ$ phase shift (AFM
domain wall) that yields the negative correlations in (c).}
\label{fig:bhe_obc} \end{figure*}

For theoretical simplicity, in the main text we have
addressed the behavior of both the TFI chain and the MITP
model with periodic boundary conditions. In the
thermodynamic limit, boundary conditions have no effect on
the bulk behavior. However, in experimental practice
\cite{Simon2011_1}, the MITP model is realized in a finite
(indeed, small) system with open boundary conditions. To
better understand the experimental consequences of our work,
we now address the KZ dynamics of the MITP model with open
boundary conditions. We solve for the spin-spin correlation
function $G(x)$ for the same criteria as in fig. 2 from the
main text, with the results show in fig. \ref{fig:bhe_obc}.

Open boundary conditions (o.b.c.) in the MITP model differ
from periodic boundary conditions (p.b.c.) only by the
absence of the projector $\mathcal{P}$ on the ends of the chain,
which in p.b.c.
projects out configurations with up spins on both the first and last
sites. As a result, on the ordered side of the phase
transition, both the first and last sites will favor spin up
over spin down due to the Zeeman field
(see fig. \ref{fig:bhe_obc}d). This results in a large
even/odd effect with regards to the system size: for odd
size systems, there is no problem creating the
staggered AFM state with up spins on each end, but for even size systems this
ordering becomes frustrated. Note that a similar effect
can be seen with p.b.c.: for $L$ odd the AFM is
frustrated.

An additional complication with o.b.c. is that, compared to
the translationally invariant system with p.b.c., there
is no unique way to define $G(x)$;
due to the loss of translation invariance, \begin{equation}
\label{eq:Gdef} \langle S_j^z S_{j+x}^z \rangle-\langle
S_j^z\rangle\langle S_{j+x}^z \rangle \neq \langle S_j^z
S_{j+x}^z \rangle-\langle S^z\rangle^2 ~, \end{equation}
where $\langle S^z \rangle \equiv \frac{1}{L}\sum_j \langle
S_j^z \rangle $. Choosing either side of (\ref{eq:Gdef}) in
defining $G(x)$ gives quantitatively different results.
Similarly, including sites near the end of the chain in the definition
of $G(x)$ introduces end effects whereas considering only spin
correlations between sites in the middle third of the chain
emphasizes the bulk behavior
at the expense of throwing away data.
The results we show in this supplement use the definition
\begin{equation}
G(x) \equiv
(-1)^x\frac{1}{L-x}\sum_{j=1}^{L-x}\left[ \langle S_j^z
S_{j+x}^z \rangle- \langle S^z \rangle^2 \right] ~.
\end{equation}
Empirically (data not shown), the various definitions of
$G(x)$ that we described above gave quantitatively
different, but qualitatively similar, results, that all
approached the p.b.c. solution -- as discussed below -- in a
similar manner.

The spin-spin correlations resulting from a KZ ramp of the
MITP with o.b.c. are shown in fig. \ref{fig:bhe_obc}, panels
a-c. The even/odd effects discussed above are apparent in a
slow sweep (fig. \ref{fig:bhe_obc}a), with neither even nor
odd size systems matching the p.b.c. solution. In general,
matching between o.b.c. and p.b.c. solutions improves in the
limit of fast sweeps (fig. \ref{fig:bhe_obc}b,c). This is
not surprising, as finite size effects vanish in the KZ-TDL,
so boundary conditions should have no effect in that limit.
Note that fig. \ref{fig:bhe_obc} shows scaled correlations
as a function of scaled length $x/L$, and we see
that the solutions match better at small
distances $x\ll L$, where boundary effects are not
important, as opposed to $x/L \sim 0.5$, where boundary
effects are non-trivial.

Most importantly, the athermal negative correlations that we
saw with p.b.c. exist with o.b.c. (fig. \ref{fig:bhe_obc}c),
and therefore are robust with respect to both boundary conditions
and small system size. In fact, o.b.c. enhances the ability
to see negative correlations
experimentally. As seen in fig. \ref{fig:bhe_obc}d, due to
the absence of the end projector, the system breaks the
$\mathbb{Z}_2$ Ising symmetry and selects the AFM with
higher spin-up probability at the ends. By breaking the
degeneracy of the ground states, the domain walls associated
with negative correlations can be seen simply by measuring
the expectation value of $S_j^z$ for each site $j$, rather
than measuring correlation functions. As expected from fig.
\ref{fig:bhe_obc}c, for constant $\Lambda$ these domain
walls occur at a roughly constant value of $j/L$. This
prediction, confirmed in fig. \ref{fig:bhe_obc}d, implies
that the position of the domain wall relative to the edge of
the sample is tunable by the ramp rate $v$.

Based on these observations, we estimate the ramp rates
necessary to see athermal spin correlations at distance
$x$ equal to half the system size, using
the definition of $G(x)$ from above.  The results, shown
in the inset to Fig. 2c in the main text for a twelve-site
bosonic chain, indicate the the time scales necessary to see this
phenomenon are already within experimental reach.  However,
we should note that for the smallest system shown in
Ref.~\cite{Simon2011_1}, a six-site bosonic chain, the negative
correlations do not appear due to finite size effects.  For
any even length chain larger than six sites, these athermal correlations
do robustly appear.

In conclusion, while open boundary conditions have a
non-trivial effect on the scaling dynamics of the MITP
model, these effects are diminished in the limit $Lv^{1/2}
\gg 1$ or $x \ll L$. Crucially, the qualitatively athermal
property of negative correlations survives for o.b.c. and
small system size $L$. In fact, the breaking of the ground
state degeneracy that occurs as a result of boundary
conditions makes seeing these negative correlations easier
experimentally, as domain walls in measurements of $S^z$.
Therefore, we believe that many of the qualitative features
of the TFI scaling functions should be experimentally
accessible, while the quantitative scaling collapse will
require larger system size or some mechanism to compensate
for the open boundary conditions (c.f.e.
\cite{Kuhner2000_1}).

\end{document}